\documentclass[aps,prl,superscriptaddress,showpacs,floatfix,twocolumn,10pt,longbibliography]{revtex4-1}

\usepackage[T1]{fontenc}

\usepackage{bm}
\usepackage{amsmath}
\usepackage{amssymb}
\usepackage{latexsym}
\usepackage{amsfonts}
\usepackage{graphicx}
\usepackage{color}
\usepackage{grffile}

\usepackage[colorlinks=true,urlcolor=blue,linkcolor=blue,citecolor=blue]{hyperref}

\newcommand{\be}{\begin{equation}}
\newcommand{\ee}{\end{equation}}

\newcommand{\ben}{\begin{eqnarray}}
\newcommand{\een}{\end{eqnarray}}

\newcommand{\vrr}{{\bf{r}}}

\def\(({\left(}
\def\)){\right)}
\def\[[{\left[}
\def\]]{\right]}

\begin{document}

\title{Scaling laws for stationary Navier-Stokes-Fourier flows and \\ the unreasonable effectiveness of hydrodynamics at the molecular level}

\author{P.I. Hurtado}
\email{phurtado@onsager.ugr.es}
\affiliation{Institute Carlos I for Theoretical and Computational Physics, and Departamento de Electromagnetismo y F\'{\i}sica de la Materia, Universidad de Granada, 18071 Granada, Spain}
\author{J.J. del Pozo}
\email{j.j.d.pozo.mellado@tue.nl}
\affiliation{Eindhoven Hendrik Casimir Institute, and Department of Applied Physics and Science Education, Eindhoven University of Technology, P. O. Box 513, 5600 MB Eindhoven, The Netherlands}
\author{P.L. Garrido}
\email{garrido@onsager.ugr.es}
\affiliation{Institute Carlos I for Theoretical and Computational Physics, and Departamento de Electromagnetismo y F\'{\i}sica de la Materia, Universidad de Granada, 18071 Granada, Spain}

\date{\today}

\pacs{
}

\begin{abstract}
Hydrodynamics provides a universal description of the emergent collective dynamics of vastly different many-body systems, based solely on their symmetries and conservation laws. Here we harness this universality, encoded in the Navier-Stokes-Fourier (NSF) equations, to find general scaling laws for the stationary uniaxial solutions of the compressible NSF problem far from equilibrium. We show for general transport coefficients that the steady density and temperature fields are functions of the pressure and a kinetic field that quantifies the quadratic excess velocity relative to the ratio of heat flux and shear stress. This kinetic field obeys in turn a spatial scaling law controlled by pressure and stress, which is inherited by the stationary density and temperature fields. We develop a scaling approach to measure the associated master curves, and confirm our predictions through compelling data collapses in large-scale molecular dynamics simulations of paradigmatic model fluids. Interestingly, the robustness of the scaling laws in the face of significant finite-size effects reveals the surprising accuracy of NSF equations in describing molecular-scale stationary flows. Overall, these scaling laws provide a novel characterization of stationary states in driven fluids.
\end{abstract}

\maketitle

\emph{Introduction}.
The Navier-Stokes-Fourier (NSF) equations govern the macroscopic dynamics of compressible, viscous, and heat-conducting fluids \cite{landau13a, de-groot13a, zarate06a, feireisl04a, novotny04a, zeytounian12a}, being the cornerstone of continuum fluid dynamics across physics and engineering. They are are based on the local conservation of energy, momentum and mass density, together with the constitutive laws of Fourier and Newton \cite{landau13a, de-groot13a, zarate06a}, and can be derived from Boltzmann kinetic equation under suitable assumptions \cite{chapman90a, resibois77a}. Despite their widespread modeling success \cite{zeytounian12a} and importance for developments in pure mathematics \cite{lions96a, bresch07a, feireisl01a, feireisl07a, mucha09a, feireisl10a, mucha10a, novotny11a, feireisl12a, chaudhuri22a, pokorny22a} and numerical analysis \cite{feng15a, feireisl16a, dumbser16a, saadat19a, saadat21a}, deriving general properties for their solutions remains challenging. In particular, stationary solutions of the NSF equations --those describing steady flows maintained by external gradients-- are most relevant as they capture the macroscopic organization of matter under sustained driving, and are central to understanding transport processes, fluid stability or energy conversion, providing a natural arena to probe the structure of hydrodynamic laws beyond linear response. Yet, the structure of these solutions remains analytically elusive due to the nonlinear and coupled nature of the governing equations, particularly for compressible fluids with realistic transport coefficients.

Here we uncover a set of general scaling laws that govern the stationary solutions of the compressible NSF equations under uniaxial temperature and velocity gradients. We show that, for general nonlinear transport coefficients, the stationary density and temperature fields are local functions of a kinetic field and the fluid's pressure. This kinetic field, which measures the local quadratic excess velocity with respect to the ratio of the total heat current over the shear stress, obeys in turn a biparametric spatial scaling law controlled by the shear stress and the pressure, which is inherited by the stationary density and temperature fields. These results thus uncover a striking simplification: the macroscopic structure of arbitrary nonequilibrium steady flows collapses onto universal master curves determined solely by the transport coefficients and equation of state. We confirm these scaling laws via compelling data collapses from massive computer simulations of two molecular fluids, namely three-dimensional Lennard-Jones fluids and two-dimensional hard disks systems. The remarkable scaling observed across a wide range of gradients, pressures, and system sizes demonstrates that the stationary state of a molecular fluid is entirely specified by the flowing currents and pressure, independently of system size or any other microscopic details, and is given by the solution of the macroscopic NSF equations for these bulk flow conditions. This reveals the surprising effectiveness of NSF hydrodynamics in describing stationary nonequilibrium behavior at the molecular scale, providing a powerful framework to characterize and predict fluid flows via general scaling laws.

\emph{Scaling laws in NSF hydrodynamics.}
We consider a $d$-dimensional compressible, viscous and heat-conducting fluid in a fixed volume driven into a nonequilibrium steady state by two thermal walls at $x=0,1$ at temperatures $T_0\ge T_1$, which also move along the $y$-direction with velocities $v_0$ and $v_1$, thus imposing a combined heat and shear flow. The macroscopic structure of this steady state follows from the stationary solution of the NSF equations for the density $\rho(x)$, $y$-velocity $v(x)$ and temperature $T(x)$ fields \cite{landau13a, de-groot13a, truesdell66a, zarate06a, feireisl04a, novotny04a, zeytounian12a}, namely $\partial_x \pi(\rho,T)=0$, $\partial_x[\eta(\rho,T)\,\partial_x v(x)]=0$ and $\partial_x [\kappa(\rho,T)\,\partial_x T(x)]+\eta(\rho,T)\,[\partial_x v(x)]^2=0$. Here $\pi(\rho,T)$ is the fluid's local pressure, and $\kappa(\rho,T)$ and $\eta(\rho,T)$ are the fluid's heat conductivity and the shear viscosity, respectively, while boundary conditions are $T(x=0,1)=T_{0,1}$ and $v(x=0,1)=v_{0,1}$, which are of no-slip type for $v(x)$ \cite{chernov95a,chernov97a}. The NSF equations can be integrated to obtain
\begin{align}
\pi(\rho,T)&=P \,,\label{NSF1}\\
\eta(\rho,T)\,\partial_x v(x) &= \sigma \,,\label{NSF2}\\
-\kappa(\rho,T)\,\partial_x T(x) &= J + \sigma v(x) \,,\label{NSF3}
\end{align}
where $P$, $\sigma$ and $J$ are the fluid's pressure, shear stress and heat current, respectively. The function $\pi(\rho,T)$ in Eq.~\eqref{NSF1} is usually given by the fluid's \emph{equilibrium} equation of state (EoS), a property known as macroscopic local equilibrium which has been tested with high accuracy in many nonequilibrium fluids over a wide range of states \cite{tenenbaum82a, tenenbaum83a, bresme14a, pozo15a, pozo15b, matsubara16a, hurtado16a, lautenschlaeger19a, lautenschlaeger19b, hurtado20a}. We may now invert the EoS~\eqref{NSF1} (valid away from critical points) to write $T(x)=\tau_P[\rho(x)]$, such that $\pi[\rho,\tau_P(\rho)]=P$, so the transport coefficients are $\kappa(\rho,T)= \kappa_P(\rho)$ and $\eta(\rho,T)= \eta_P(\rho)$. Interestingly, Eq.~\eqref{NSF2} implies that for $\sigma\ne 0$ (or equivalently $v_0\ne v_1$) the velocity field will be always monotonous, so the local velocity can be used as a proxy of space. Hence, noting that $\partial_x T = \tau'_P(\rho) \partial_x\rho$, with $'$ denoting derivative with respect to the argument, and using the chain rule and Eq.~\eqref{NSF2} to write $\partial_x\rho = \partial_v\rho~\partial_x v = \sigma \partial_v \rho /\eta_P(\rho)$, we can rewrite Eq.~\eqref{NSF3} as
\be
-\frac{\kappa_P(\rho) \tau'_P(\rho)}{\eta_P(\rho)} \partial_v \rho \equiv {\cal G}'_P(\rho) \partial_v \rho = \partial_v {\cal G}_P(\rho) = v + \frac{J}{\sigma} \, ,
\label{dGdu0}
\ee
where we have defined a new function ${\cal G}'_P(\rho)$ in the first equality. This equation can be simply integrated to obtain ${\cal G}_P(\rho)= \frac{1}{2}\left(v + \frac{J}{\sigma}\right)^2 + \xi$ with $\xi$ some integration constant. In this way, the uniparametric function ${\cal G}_P[\rho(x)]$ is always a quadratic form of the local velocity $v(x)$, for any boundary driving and arbitrary transport coefficients. Defining now a \emph{kinetic field} $\omega(x)\equiv \frac{1}{2}\left(v(x) + \frac{J}{\sigma}\right)^2$ and the inverse function ${\cal R}_P(\cdot) \equiv {\cal G}_P^{-1}(\cdot)$ \footnote{We assume here that the function ${\cal G}_P(\rho)$ is invertible, which seems reasonable given that density profiles are smooth and well-behaved $\forall x$ [and hence $\forall v(x)]$.}, we find
\be
\rho(x) = {\cal R}_P\left(\omega(x)+\xi\right) \, , \quad T(x) =  {\cal T}_P(\omega(x)+\xi) \, ,
\label{rhow}
\ee
where we have defined ${\cal T}_P(\cdot)\equiv \tau_P\left[{\cal R}_P(\cdot)\right]$. Therefore, the local density and temperature fields are sole functions of pressure and the local kinetic field $\omega(x)$, that measures the quadratic excess velocity with respect to the ratio of the total heat current over the shear stress. Indeed, there exists a unique pair of master surfaces ${\cal R}_P$ and ${\cal T}_P$ in $\omega-P$ space from which any stationary density and temperature profiles follow after an appropriate shift $\xi$, when written in terms of the local kinetic field $\omega(x)$. 

The kinetic field $\omega(x)$ obeys in turn a simple spatial scaling law. To show this, we first note from the definition of $\omega(x)$ that $\partial_x v = \pm (2\omega)^{-1/2} \partial_x w$, so Newton's law~\eqref{NSF2} can be now written as $(2\omega)^{-1/2} \eta_P[{\cal R}_P(\omega+\xi)]\partial_x \omega \equiv {\cal H}'_{P,\xi}(\omega) \partial_x \omega = \partial_x {\cal H}_{P,\xi}(\omega) = \pm \sigma$ after introducing an additional function ${\cal H}'_{P,\xi}(\omega)$. Therefore ${\cal H}_{P,\xi}[\omega(x)]$ is a linear function of space, ${\cal H}_{P,\xi}(\omega) = \pm\sigma x +\zeta$, with $\zeta$ an additional integration constant. Defining now the inverse biparametric function ${\cal W}_{P,\xi}(\cdot) \equiv {\cal H}_{P,\xi}^{-1}(\cdot)$ \footnote{Since we expect smooth velocity profiles $v(x)$, and hence well-defined kinetic fields $\omega(x)$, we anticipate the new biparametric function ${\cal H}_{P,\xi}(\omega)$ to be invertible.}, we get
\be
\omega(x) = {\cal W}_{P,\xi}(\pm\sigma x +\zeta) \, .
\label{H2}
\ee
Therefore there exists a parametric family of master surfaces ${\cal W}_{P,\xi}$ in $x-\xi$ space, one for each pressure $P$, from which all kinetic field profiles $\omega(x)$ follow after scaling space by the measured shear stress $\sigma$ and an appropriate constant shift $\zeta$. Remarkably, the density and temperature fields automatically inherit this spatial scaling via their simple dependence on the kinetic field, see Eq.~\eqref{rhow}. In particular, $\rho(x) = {\cal R}_P\left[{\cal W}_{P,\xi}(\pm\sigma x +\zeta) +\xi \right]$ and $T(x) = {\cal T}_P\left[{\cal W}_{P,\xi}(\pm\sigma x +\zeta) +\xi  \right]$, and therefore
\be
\rho(x) = \bar{\cal R}_{P}(\pm\sigma x +\zeta,\omega+\xi) \, , \quad T(x)  = \bar{\cal T}_{P}(\pm\sigma x +\zeta,\omega+\xi) \, ,
\label{rhoTx}
\ee
with $\bar{\cal R}_{P}(\cdot,\cdot)$ and $\bar{\cal T}_{P}(\cdot,\cdot)$ two new parametric surfaces of the scaled space and the shifted kinetic field. Remarkably, the scaling laws~\eqref{rhow}-\eqref{rhoTx} are valid for arbitrary boundary driving and any type of fluid, with the shape of the scaling functions depending exclusively on the fluid's viscosity, heat conductivity and EoS.

Systems with hard-particle interactions \cite{mulero08a} (as hard disks) exhibit density-temperature separability for both the EoS, $P=T\, \pi(\rho)$, and the transport coefficients, $\kappa(\rho,T)=\sqrt{T}\, \kappa(\rho)$ and $\eta(\rho,T)=\sqrt{T}\, \eta(\rho)$ \cite{pozo15b,gnan09a}. This simplifies the scaling laws derived above by making explicit the pressure dependence, i.e. $\tau_P(\rho) = P/\pi(\rho)$ while $\kappa_P(\rho)=\sqrt{P}\, \kappa(\rho)/\sqrt{\pi(\rho)}$ and $\eta_P(\rho)=\sqrt{P}\, \eta(\rho)/\sqrt{\pi(\rho)}$, which results in the definition ${\cal G}'_P(\rho)=P\, \kappa(\rho)\pi'(\rho)/[\eta(\rho)\pi(\rho)^2]$, see Eq.~\eqref{dGdu0}. We thus obtain simpler scaling laws based on a pair of master \emph{curves} 
\be
\rho(x) = {\cal R}\left(\frac{\omega(x)}{P} + \xi \right) , \quad T(x) = P\, {\cal T}\left(\frac{\omega(x)}{P} + \xi \right) \, .
\label{rhoThd}
\ee
Moreover, the kinetic field now scales as $\omega(x) = P\, {\cal W}_\xi\left(\pm \frac{\sigma}{P}x + \zeta \right)$, where ${\cal W}_\xi(\cdot)\equiv {\cal H}_{\xi}^{-1}(\cdot)$ is now a \emph{unique} surface in $x-\xi$ space, obtained by integrating and inverting the definition ${\cal H}'_{\xi}(\omega/P)\equiv \left(2(\omega/P) \, \pi[{\cal R}(\omega/P+\xi)]]^{-1/2} \eta[{\cal R}(\omega/P+\xi)\right)$. The density and temperature fields also inherit this spatial scaling, see Eq.~\eqref{rhoThd}, so $\rho(x) = \bar{\cal R}\left(\pm \frac{\sigma}{P}x + \zeta, \frac{\omega}{P}+\xi \right)$ and $T(x) = P\, \bar{\cal T}\left(\pm \frac{\sigma}{P}x + \zeta, \frac{\omega}{P}+\xi  \right)$. This simplified scaling laws can be generalized for the broad family of inverse-power-law fluids, characterized by homogeneous interparticle potentials $V(\vrr)\propto r^{-n}$, which also exhibit density-temperature separability \cite{pozo15b, gnan09a}.

\emph{Scaling in molecular fluids}.
The previous scaling laws apply to the macroscopic NSF equations for continuous flows. A natural question is their validity for molecular systems, where fluctuations and finite-size effects are dominant. We hence performed massive molecular dynamics simulations of two paradigmatic model fluids, namely two-dimensional hard disk ($\text{HD}_{2d}$) systems \cite{mulero08a, rapaport09a, isobe16a} and three-dimensional Lennard-Jones ($\text{LJ}_{3d}$) fluids \cite{frenkel23a, allen17a, evans14a, rapaport04a}. These particle systems are among the most inspiring, successful and prolific models of physics, as they contain the main ingredients to capture a broad class of emergent phenomena \cite{mulero08a, rapaport09a, isobe16a, frenkel23a, allen17a, evans14a, rapaport04a, chaikin00a, cates00a, berthier11a, degennes93a, mehta07a, alder59a, alder62a, alder67a, alder70a, resibois77a, rosenbluth54a, bonetto00a, dhar01a, garrido01a, savin02a, grassberger02b, narayan02a, hurtado06a, hurtado11b, chernov95a, chernov97a, bonetto97a, garrido04a, szasz00a, hurtado16a, hurtado20a, garrido22a, kob95a, kob97a, lacevic03a, trudu06a, hurtado07a, chaudhuri07a, chaudhuri10a, testard11a, chaudhuri15a, puertas03a, matsubara16a, rosales-pelaez19a, lautenschlaeger19b, stephan20a}. Both models, described in detail in the Appendix, consist in $N$ particles in a $d$-dimensional box at a given volume fraction $\phi$, and driven out of equilibrium by stochastic boundary walls \cite{livi17a, dhar08a, lepri03a, bonetto00a} at $x=0,1$ characterized by boundary temperatures $T_{0,1}$ and average $y$-velocities $v_{0,1}$ \cite{chernov95a, chernov97a} (periodic conditions are imposed along all other directions). For $T_0\ne T_1$ and $v_0\ne v_1$, net currents of energy and $y$-momentum along the $x$-direction appear driving the fluid to a nonequilibrium steady state. We measured the local hydrodynamic fields along the gradient direction, as well as $P$, $J$ and $\sigma$ using different methods, for a broad set of system sizes, volume fractions and boundary drivings. Hereafter all observables are shown in reduced units (denoted with $^*$, see Appendix) \cite{lautenschlaeger19a, lautenschlaeger19b}. 

\begin{figure}[t]
\includegraphics[width=8.5cm]{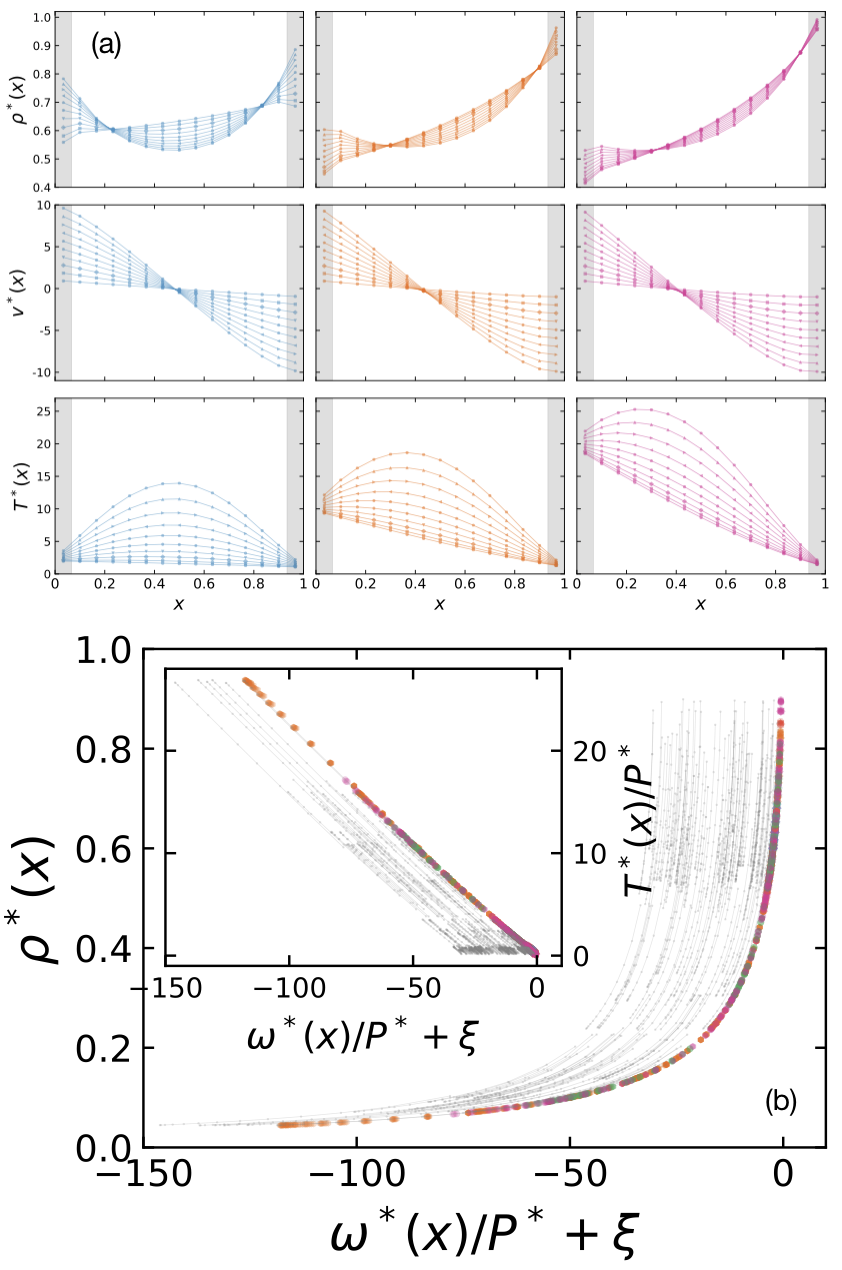}
\caption{\small (Color online) NSF scaling for $\text{HD}_{2d}$ fluids. (a) Sample of measured average profiles for the density (top row), $y$-velocity (middle row) and temperature (bottom row) for $N=1927$, $\phi=0.5$, $T_1^*=1$ and (from left to right), $T_0^*=2,~10,~20$. Each plot shows profiles for 10 different values of $v_0^*=-v_1^*\in[1,10]$. Colors codify $T_0^*$, while symbols represent different $v_0^*$. Gray bands near $x=0,1$ signal the boundary layers. (b) Scaling plot of the density $\rho^*(x)$ as a function of the reduced kinetic field $\omega^*(x)/P^*$ before (light gray) and after (color symbols) the shifts $\xi$. Inset: The same but for the reduced temperature $T^*(x)/P^*$. All data for different $N$, $\phi$, $T_0^*$ and $v_0^*$ collapse on a pair of master curves, as predicted by the NSF scaling laws~\eqref{rhoThd} for $\text{HD}_{2d}$. 
}
\label{figHD1a}
\end{figure}

\begin{figure*}[t]
\includegraphics[width=18cm]{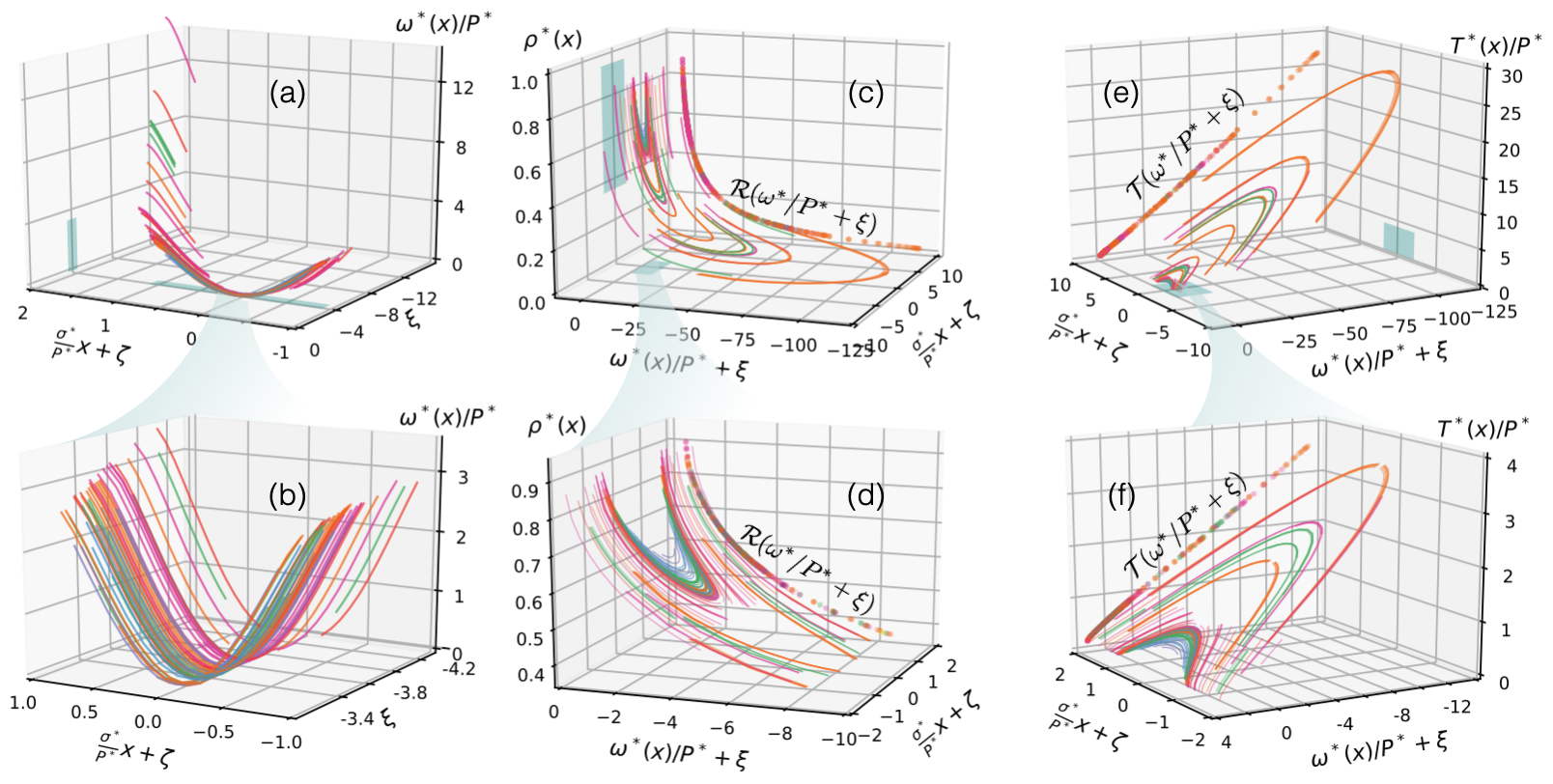}
\caption{\small (Color online) Spatial scaling for $\text{HD}_{2d}$ fluids. (a)-(b) Scaling plot of the reduced kinetic field $\omega^*(x)/P^*$ as a function of the scaled spatial variable $\frac{\sigma^*}{P^*}x$ and the shifts $\xi$ obtained from the density (and temperature) scaling, see Fig.~\ref{figHD1a}. All kinetic field profiles collapse onto a universal master surface ${\cal W}_\xi(\pm \frac{\sigma^*}{P^*}x+\zeta)$ after an appropriate spatial shift $\zeta$. This spatial scaling is inherited by the density [(c)-(d)] and reduced temperature [(e)-(f)] fields, which collapse onto two master surfaces $\bar{\cal R}\left(\pm \frac{\sigma}{P}x + \zeta, \frac{\omega}{P}+\xi \right)$ and $\bar{\cal T}\left(\pm \frac{\sigma}{P}x + \zeta, \frac{\omega}{P}+\xi  \right)$, respectively, as predicted by the NSF scaling laws. Panels (b), (d) and (f) are zooms over the corresponding shaded areas.
}
\label{figHD2}
\end{figure*}

Fig.~\ref{figHD1a}.a shows a sample of the measured hydrodynamic profiles for $\text{HD}_{2d}$ systems. In all cases the profiles obtained are  nonlinear and exhibit finite-size effects with varying $N\in[1927,8838]$. However, the local values of $\rho^*(x)$ and $T^*(x)$ in each case are tightly related by the equilibrium EoS for the observed finite-sized pressure $P^*$ (not shown), thus confirming the macroscopic local equilibrium property in this flow situation \cite{pozo15a}. Note that the thermal walls at $x=0,1$ perturb the structure of the nearby fluid, affecting two boundary layers where finite-size corrections accumulate, see gray regions near $x=0,1$ in Fig~\ref{figHD1a}.a. We neglect data from these boundary layers (just one cell near each thermal wall) to analyze the scaling behavior of the remaining bulk profiles. 

To obtain the master curve ${\cal R}(\cdot)$ of Eq.~\eqref{rhoThd} for $\text{HD}_{2d}$ fluids, we plot the measured bulk density profiles $\rho^*(x)$ as a function of the reduced kinetic field $\omega^*(x)/P^*=\frac{1}{2 P^*}(v^*(x) + \frac{J^*}{\sigma^*})^2$, obtained from the measured local velocity field $v^*(x)$, pressure $P^*$, heat current $J^*$ and shear stress $\sigma^*$ in each case (all exhibiting finite-size corrections), see thin gray curves in the main plot of Fig.~\ref{figHD1a}.b. Each curve is then shifted by a constant $\xi$ along the abscissa to yield the best collapse among all scaled profiles. The optimal shifts are obtained via minimization of a standard collapse metric \cite{bhattacharjee01a, hurtado16a} measuring the relative average distance among all pairs of overlapping curves. Using the same set of shifts, we obtain the collapse of the reduced bulk temperature profiles $T^*(x)/P^*$ as a function of $\omega^*(x)/P^*+\xi$, see inset in Fig.~\ref{figHD1a}.b. The precision of the resulting collapses is striking; a total of 2431 data points in $\rho^*-T^*-\omega^*$ space for widely different $N$, $\phi$, and gradients $\Delta T^*=T^*_0-T^*_1$ and $\Delta v^*=v^*_0-v^*_1$ collapse onto a pair of master curves as predicted by NSF hydrodynamics, see Eq.~\eqref{rhoThd}, with no appreciable corrections even though the measured data all exhibit finite-size effects. This means that the bulk of a molecular fluid (with as few as e.g.$N\lesssim 2000$ particles) self-organizes into a macroscopic flow state fully described by NSF hydrodynamics equations for the corresponding (finite-size) heat current, shear stress and pressure imposed by the external boundary driving.

In addition, we can collapse now all reduced kinetic field profiles $\omega^*(x)/P^*$ onto the master surface ${\cal W}_{\xi}(\cdot)$ by scaling space $x\to \sigma^* x/P^*$ using the measured shear stress and pressure in each case, and shifting in space the resulting reduced kinetic field profile a constant $\zeta$ so as to minimize the distance \cite{bhattacharjee01a, hurtado16a} with neighboring curves along the $\xi$-axis. Figs.~\ref{figHD2}.a-b show the master surface $\mathcal{W}_\xi(\frac{\sigma^*}{P^*}x+\zeta)$ measured in this way for the $\text{HD}_{2d}$ fluid, and the resulting collapse is excellent. This spatial scaling is inherited by the density and temperature fields, see Figs.~\ref{figHD2}.c-f, which collapse onto two universal master surfaces $\bar{\cal R}\left(\pm \frac{\sigma}{P}x + \zeta, \frac{\omega}{P}+\xi \right)$ and $\bar{\cal T}\left(\pm \frac{\sigma}{P}x + \zeta, \frac{\omega}{P}+\xi  \right)$, respectively, as predicted by the NSF scaling laws. 

For $\text{LJ}_{3d}$ fluids there is no density-temperature separability, and the predicted master functions depend parametrically on pressure. To test the scaling theory and measure the master functions, we implement a two-stage simulation protocol to generate $\text{LJ}_{3d}$ data under strictly isobaric (equal $P^*$) conditions. This involves an initial iterative adjustment of the system volume for each set of parameters ($N$, $\phi$, $T^*_{0,1}$, $v^*_{0,1}$) to reach the target global pressure, followed by production runs to measure the steady-state hydrodynamic profiles in each case, see Appendix for details. Using this method we collected 3120 high-quality data points in $\rho^*-T^*-\omega^*$ space for $N\in[10^3,10^4]$ and 9 different pressures (28080 data points in total), enabling a consistent measurement of the master functions for $\text{LJ}_{3d}$ systems across a broad range of parameters. Indeed, using the same scaling technique as above for each $P^*$, we obtained excellent data collapses (with no finite-size corrections) for the master functions ${\cal R}_P(\cdot)$, ${\cal T}_P(\cdot)$ and ${\cal W}_{P,\xi}(\cdot)$, which display a non-trivial dependence on pressure, see figures in Appendix.

{\emph{Discussion}}. 
Our results reveal a profound simplification in the structure of stationary solutions of the compressible Navier-Stokes-Fourier equations. By identifying a set of scaling laws that govern nonequilibrium steady uniaxial flows, we demonstrate that the stationary hydrodynamic fields --typically obtained by solving a highly nonlinear, coupled set of partial differential equations-- can instead be reconstructed algebraically from a small set of bulk parameters, namely the heat current $J$, shear stress $\sigma$, and pressure $P$, along with two scalars ($\xi$ and $\zeta$), which encapsulate the particular steady-state realization of the system. The entire spatial structure of the stationary flow then follows from these bulk constants and some master curves, built from the fluid's transport coefficients and equation of state. In this way, a few emergent bulk quantities fully encapsulate the steady-state response of the system. Most remarkably, this behavior holds not only in the hydrodynamic limit but extends down to molecular fluids of modest size, which exhibit steady profiles in precise agreement with the NSF-based scaling laws. This robustness of the scaling laws under finite-size effects demonstrates that the bulk structure of the molecular fluid is fully encoded in a number of bulk invariants ($J$, $\sigma$, $P$, $\xi$ and $\zeta$), independently of how these quantities are established microscopically via boundary conditions. This provides not only a powerful tool for the analysis and prediction of nonequilibrium flows, but also a conceptual advance in how we understand steady-state hydrodynamics: as a universal theory governed by master curves and bulk invariants, rather than boundary-value solutions of differential equations. It opens the door to novel theoretical and computational approaches to fluid transport, and offers compelling evidence for the unreasonable effectiveness of continuum hydrodynamics in describing driven molecular fluids far beyond its traditional domain of applicability.

\acknowledgements

The research leading to these results has received funding from the I+D+i grants PID2023-149365NB-I00, PID2020-113681GB-I00, and C-EXP-251-UGR23, funded by MICIU/AEI/10.13039/501100011033/, ERDF/EU, and Junta de Andaluc\'{\i}a - Consejer\'{\i}a de Econom\'{\i}a y Conocimiento. We are also grateful for the the computing resources and technical support provided by PROTEUS, the supercomputing center of Institute Carlos I for Theoretical and Computational Physics in Granada, Spain.

\appendix

\section{Appendix}

In order to test the scaling laws derived for NSF hydrodynamics, we performed massive molecular dynamics simulations of two paradigmatic model fluids, namely two-dimensional hard disk ($\text{HD}_{2d}$) systems \cite{mulero08a, rapaport09a, isobe16a} and three-dimensional Lennard-Jones ($\text{LJ}_{3d}$) fluids \cite{frenkel23a, allen17a, evans14a, rapaport04a}. These particle systems represent some of the most influential and productive models in physics, as they encapsulate the essential features required to understand a wide spectrum of emergent phenomena \cite{mulero08a, rapaport09a, isobe16a, frenkel23a, allen17a, evans14a, rapaport04a, chaikin00a, cates00a, berthier11a, degennes93a, mehta07a, alder59a, alder62a, alder67a, alder70a, resibois77a, rosenbluth54a, bonetto00a, dhar01a, garrido01a, savin02a, grassberger02b, narayan02a, hurtado06a, hurtado11b, chernov95a, chernov97a, bonetto97a, garrido04a, szasz00a, hurtado16a, hurtado20a, garrido22a, kob95a, kob97a, lacevic03a, trudu06a, hurtado07a, chaudhuri07a, chaudhuri10a, testard11a, chaudhuri15a, puertas03a, matsubara16a, rosales-pelaez19a, lautenschlaeger19b, stephan20a}. 

Both models consist in $N$ particles of mass $m$ and diameter $\ell$ in a $d$-dimensional box of lengths $L_\parallel$ and $L_\perp\le L_\parallel$ along the gradient ($x$) and all other orthogonal directions, respectively. To investigate the large-$N$ limit at constant, non-zero boundary gradients, we fix $L_\parallel=1$ so the particle diameter for a given volume fraction $\phi$ depends on $N$ as $\ell(N)=2[\phi L_\perp^{d-1} \Gamma(d/2+1)/(N \pi^{d/2})]^{1/d}$. The gradients driving the system out of equilibrium are imposed via two stochastic boundary walls \cite{livi17a, dhar08a, lepri03a, bonetto00a} at $x=0,1$ characterized by a $d$-dimensional Maxwellian velocity distribution with temperature $T_{0,1}$ and average $y$-velocity $v_{0,1}$ \cite{chernov95a, chernov97a}. Moreover, periodic conditions are imposed along all other directions. $\text{HD}_{2d}$ systems feature hard-type particle interactions, for which event-driven molecular dynamics simulations were used \cite{mulero08a, rapaport09a, isobe16a}. $\text{LJ}_{3d}$ fluids are characterized instead by a soft 6-12 potential with length and energy scales $\ell$ and $\varepsilon$, respectively, which was truncated and shifted linearly  at $\ell_c=2.5\ell$ so as to guarantee a continuous potential and force at the cutoff distance $\ell_c$ \cite{frenkel23a, allen17a}. A velocity-Verlet scheme was used to integrate the equations of motion in this case.

For $T_0\ne T_1$ and $v_0\ne v_1$, net currents of energy and $y$-momentum along the $x$-direction appear driving the fluid to a nonequilibrium steady state. We measured the local density, $y$-velocity and temperature fields along the gradient direction, as well as the virial pressure, heat current and shear stress profiles, together with their wall counterparts, which coincide with the average bulk behavior \cite{pozo15b}. All observables are shown in reduced units \cite{lautenschlaeger19a, lautenschlaeger19b}, e.g. $\rho^*=\rho \ell^d$, $v^*=v \sqrt{\varepsilon/m}$, $T^*= k_\textrm{B} T /\varepsilon$, with $k_\textrm{B}$ the Boltzmann constant, while $P^*=P \ell^d/\varepsilon$, $\sigma^*=\sigma \ell^d/\varepsilon$ and $J^*=J \ell^d\sqrt{m}/\varepsilon^{3/2}$ (we choose $\varepsilon=k_\text{B} T_1$ for $\text{HD}_{2d}$ fluids). For both $\text{HD}_{2d}$ and $\text{LJ}_{3d}$ fluids we explore a broad set of system sizes and boundary drivings. To measure the different hydrodynamic profiles, we divided the simulation box into $n$ virtual cells along the gradient direction ($n=30$ for $\text{LJ}_{3d}$ and $n=15$ for $\text{HD}_{2d}$). Time averages were performed with measurements every $10$ collision per particle (cpp) on average for a total time of $10^6-10^7$ cpp, after a relaxation time of $10^3$ cpp which was found sufficient to reach the steady state. Errors are computed with 99.7\% confidence level, and errorbars in plots are smaller than the symbol sizes.

$\text{HD}_{2d}$ systems were simulated in a square box ($L_\perp=1=L_\parallel$) for 7 different $\phi\in[0.05,0.5]$, 8 different $N\in[1927,8838]$, 6 values of $T^*_0\in[1,20]$ and 10 values of $v^*_0\in [1,10]$ with $v^*_1=-v^*_0$. The results obtained for the bare hydrodynamic profiles and the measured master curves are shown and discussed in the main text.

\begin{figure}
\includegraphics[width=8.5cm]{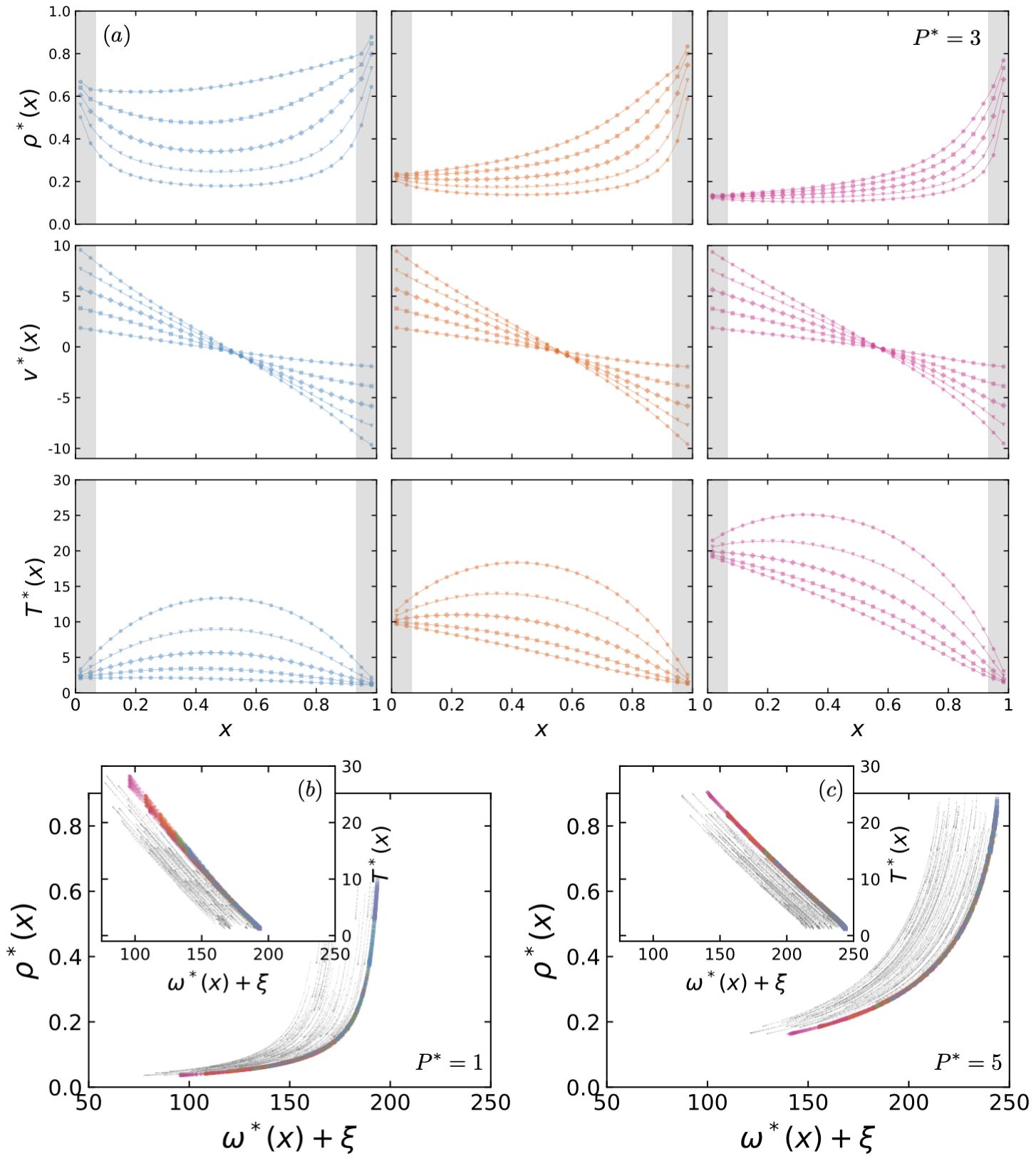}
\caption{\small (Color online) NSF scaling for $\text{LJ}_{3d}$ fluids. (a) Sample of density (top row), $y$-velocity (middle row) and temperature (bottom row) profiles measured for $N=10^4$ LJ particles, $T_1^*=1$, target pressure $P^*=3$ and (from left to right), $T_0^*=2,~10,~20$. Each plot shows profiles for 5 different values of $v_0^*=-v_1^*\in[2,10]$. Colors codify $T_0^*$, while symbols represent different $v_0^*$. Gray bands near $x=0,1$ signal the boundary layers. Bottom panels: Scaling plot of $\rho^*(x)$ vs $\omega^*(x)$ for pressures (b) $P^*=1$ and (c) $P^*=5$, before (light gray) and after (color symbols) the shifts $\xi$ along the abscissa. The insets show the equivalent temperature scalings. For each $P^*$, all data for different $N$, $T_0^*$ and $v_0^*$ collapse on a pair of master curves, as predicted by the NSF scaling laws~\eqref{rhow} for $\text{LJ}_{3d}$. The resulting master function ${\cal R}_P(\omega+\xi)$ and ${\cal T}_P(\omega+\xi)$ depend nontrivially on pressure.
}
\label{figLJ1}
\end{figure}

On the other hand, for $\text{LJ}_{3d}$ fluids there is no density-temperature separability, and the predicted scaling functions ${\cal R}_P(\cdot)$ and ${\cal T}_P(\cdot)$ depend parametrically on pressure. To measure these master functions we hence need to collapse density and temperature profiles measured under isobaric (equal $P^*$) conditions. Our simulations for given $N$, $T^*_{0,1}$ and $v^*_{0,1}$ hence proceed in two steps. First we conduct a preparation run measuring the bulk-averaged steady-state virial pressure $P^*$ while adjusting iteratively the system volume (via $L_\perp$) until the target pressure is reached with a tolerance of 1\%. Once the appropriate volume fraction $\phi$ for a given $N$, $T^*_{0,1}$, $v^*_{0,1}$ and target $P^*$ is determined, full nonequilibrium simulations are carried out to measure the steady-state hydrodynamic profiles. In total, we sampled 9 distinct pressures in the range $P^* \in [1, 5]$, as well as 4 different $N\in[10^3,10^4]$, 6 values of $T^*_0\in[1,20]$ and 5 values of $v^*_0\in [2,10]$ with $v^*_1=-v^*_0$ in an elongated $3d$ box with $L_\perp \approx 1/4$ (recall $L_\parallel=1$), enabling a consistent measurement of the scaling functions for $\text{LJ}_{3d}$ systems across a wide range of parameters. 

\begin{figure}
\includegraphics[width=9cm]{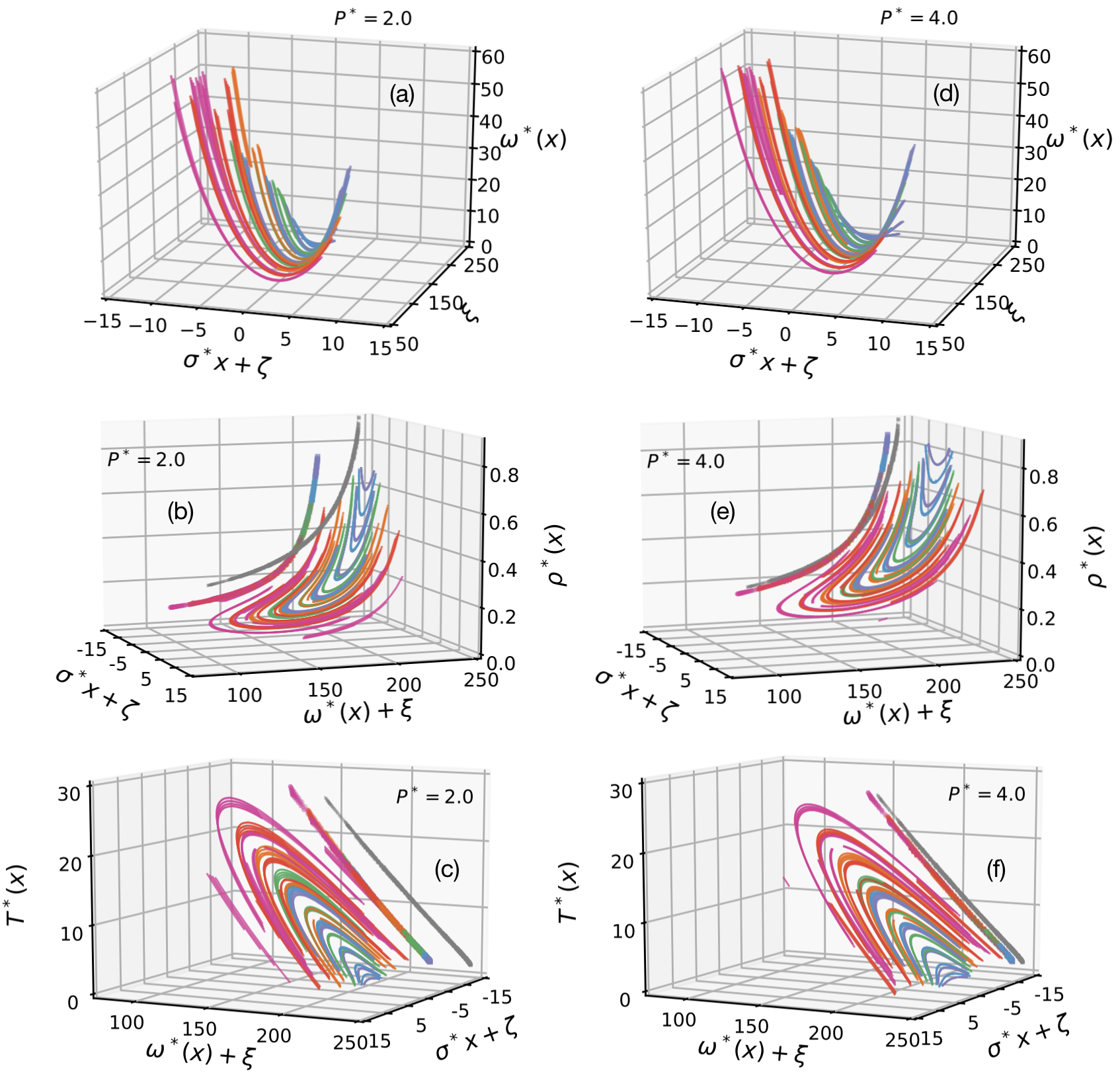}
\caption{\small (Color online) Spatial scaling for $\text{LJ}_{3d}$ fluids. Top row: Scaling plot of $\omega^*(x)$ for different $P^*$ as a function of $\sigma^* x$ and the shifts $\xi$ obtained from density (and temperature) scalings, see Fig.~\ref{figLJ1}. All kinetic field profiles collapse for each pressure onto a universal master surface ${\cal W}_{P,\xi}(\pm \sigma x+\zeta)$ after an appropriate spatial shift $\zeta$. This spatial scaling is inherited by the density (middle row) and temperature (bottom row) fields, which collapse onto two universal master surfaces $\bar{\cal R}_P\left(\pm \sigma x + \zeta, \omega+\xi \right)$ and $\bar{\cal T}_P\left(\pm \sigma x + \zeta, \omega+\xi  \right)$, respectively, as predicted by the NSF scaling laws~\eqref{rhow}. Left column [(a)-(c)] corresponds to $P^*=2$, and right column  [(d)-(f)] to $P^*=4$. Gray curves projected in the $\rho^*-\omega^*$ and $T^*-\omega^*$ planes of middle and bottom panels correspond to the $P^*=5$ scaling curves, for comparison.
}
\label{figLJ2}
\end{figure}

Fig.~\ref{figLJ1}.a shows a sample of the measured hydrodynamic profiles for $N=10^4$ and $P^*=3$. As for $\text{HD}_{2d}$ fluids, the structural perturbation of the thermal walls define two boundary layers (gray bands in Fig.~\ref{figLJ1}.a, now up to two cells near each boundary) which are discarded to analyze the scaling physics of the remaining bulk profiles. We obtain the master function ${\cal R}_P(\cdot)$ for each pressure by plotting $\rho^*(x)$ as a function of the measured local kinetic field $\omega^*(x)$ in each case, and shifting the different curves along the abscissa a constant $\xi$ to yield the best collapse according to a standard distance metric \cite{bhattacharjee01a, hurtado16a}. The same set of shifts $\xi$ allows to collapse the temperature profiles. Figs.~\ref{figLJ1}.b,c show the resulting master curves ${\cal R}_P(\cdot)$ and ${\cal T}_P(\cdot)$ measured for two different pressures and the whole set of values for $N$, $T^*_0$ and $v^*_0$, which result in 3120 data points in $\rho^*-T^*-\omega^*$ space for each $P^*$ (or 28080 data points in total). The observed collapse is excellent $\forall P^*$, as predicted by the continuous NSF scaling laws and despite finite-size effects in measurements, and the resulting master curves ${\cal R}_P(\cdot)$ and ${\cal T}_P(\cdot)$ display a non-trivial dependence on pressure. The measured kinetic field profiles $\omega^*(x)$ also collapse onto a spatial master surface ${\cal W}_{P^*,\xi}(\sigma^* x + \zeta)$ for each $P^*$ after scaling space by the observed shear stress and shifting the resulting curve in space a distance $\zeta$ to minimize the distance with neighboring curves along the $\xi$-axis, see Figs.~\ref{figLJ2}.a,d. Moreover, as dictated by the NSF scaling laws~\eqref{rhoTx}, the other hydrodynamic fields inherit this spatial scaling, so all density and temperature profiles measured for a given pressure collapse on a pair of master surfaces $\bar{\cal R}_{P}(\pm\sigma x +\zeta,\omega+\xi)$ and $\bar{\cal T}_{P}(\pm\sigma x +\zeta,\omega+\xi)$, see Figs.~\ref{figLJ2}.b,e and \ref{figLJ2}.c,f, respectively. This scaling behavior extends to all target pressures $P^* \in [1, 5]$ measured.

Ultimately, these findings strengthen the universality of the scaling laws reported in the main text. Whether in $2d$ with hard collisions, or in $3d$ with realistic Lennard-Jones particles featuring a soft core and attractive interactions, the fluids' nonequilibrium stationary states are organized in terms of a set of master functions and bulk constants. This invariance confirms that the algebraic reconstruction of steady-state profiles is not an artifact of simplified collision rules or low dimensionality, but a robust feature of hydrodynamics that holds remarkably well for complex interacting fluids even at the molecular scale.

\bibliography{/Users/phurtado/PAPERS/BIBLIOGRAPHY/referencias-BibDesk-OK}

\end{document}